# Study of isoscaling with statistical multifragmentation models

M.B. Tsang, C.K. Gelbke, X.D. Liu, W.G. Lynch, W.P. Tan,

G. Verde, H.S. Xu[*]

*National Superconducting Cyclotron Laboratory and Department of Physics and Astronomy,*

*Michigan State University, East Lansing, MI 48824, USA*

W. A. Friedman,

*Department of Physics, University of Wisconsin, Madison, WI 53706*

R. Donangelo, S. R. Souza,

*Instituto de Física, Universidade Federal do Rio de Janeiro,*

*Cidade Universitária, CP 68528, 21945-970 Rio de Janeiro, Brazil*

C.B. Das, S. Das Gupta, D. Zhabinsky[%]

*Physics Department, McGill University, 3600 University Street, Montreal, Canada H3A 2T8,*

## Abstract

Different statistical multifragmentation models have been used to study isoscaling, i.e. the factorization of the isotope ratios from two reactions, into fugacity terms of proton and neutron number, $R_{21}(N,Z)=Y_2(N,Z)/Y_1(N,Z)=C \cdot \exp(\alpha N+\beta Z)$. Even though the primary isotope distributions are quite different from the final distributions due to evaporation from the excited fragments, the values of $\alpha$ and $\beta$ are not much affected by sequential decays. $\alpha$ is shown to be mainly sensitive to the proton to neutron composition of the emitting source and may be used to study isospin-dependent properties in nuclear collisions such as the symmetry energy in the equation of state of asymmetric nuclear matter.

[*] On leave from the Institute of Modern Physics, Lanzhou, China.
[%] Research Experience for Undergraduates at Michigan State University, 2000.

**I. Introduction**

Our understanding of nuclear collision mechanisms is obtained from measuring particles emitted during nuclear collisions[1]. The importance of the isotopic degree of freedom to obtain information about charge equilibration and the charge asymmetry dependent terms of the nuclear equation-of-state has prompted recent measurements of isotope distributions beyond Z=2 [2-6]. The availability of these data makes it possible to examine systematic trends exihibited by the isotope distributions [7].

Ideally, primary fragments should be detected right after emission in order to extract information about the collisions. However, the time scale of a nuclear reaction ($10^{-20}$ s [5-6]) is much shorter than the time scale for particle detection ($\geq 10^{-10}$ s) and most particles decay to stable isotopes in their ground states before being detected. It is therefore important to study model predictions of both primary and secondary isotope distributions [8].

Recently, isotope yields from the central collisions of $^{112}$Sn+$^{112}$Sn, $^{112}$Sn+$^{124}$Sn, $^{124}$Sn+$^{112}$Sn and $^{124}$Sn+$^{124}$Sn collisions have been measured [2]. The ratio of isotope yields from two different reactions, 1 and 2, $R_{21}(N, Z) = Y_2(N, Z)/Y_1(N, Z)$, is found to exhibit an exponential relationship as a function of the isotope neutron number N, and proton number, Z [2,7].

$$R_{21}(N,Z) = Y_2(N,Z)/Y_1(N,Z) = C \cdot \exp(\alpha N + \beta Z), \qquad (1)$$

where C is the normalization factor, $\alpha$ and $\beta$ are empirical parameters.

Equation (1) can be derived from the primary isotope yields assuming that at breakup the system may be approximated by an infinite equilibrated system and employing the Grand Canonical Ensemble. In this case, predictions for the observed isotopic yield are governed by both the neutron and proton chemical potentials, $\mu_n$ and $\mu_p$ and the temperature T, plus the individual binding energies, B(N,Z), of the various isotopes [9,10].

$$Y(N,Z)=F(N,Z,T)\exp(B(N,Z)/T)\exp(N\mu_n/T+ Z\mu_p/T) \qquad (2)$$

The factor F(N,Z,T) includes information about the secondary decay from both particle stable and particle unstable states to the final ground state yields. If the main difference between system 1 and 2 is the isospin [2,9,10], then the binding energy terms in Eq. (2) cancel out in the ratio of $Y_2(N,Z)/Y_1(N,Z)$. If one further assumes that the influence of secondary decay on the yield of a specific isotope is similar for the two reactions, i.e. $F_1(N,Z,T) \approx$

$F_2(N,Z,T)$, then Equation (1) is obtained, and $\hat{\rho}_n = \exp(\Delta\mu_n/T) = \exp(\alpha)$ and $\hat{\rho}_p = \exp(\Delta\mu_p/T) = \exp(\beta)$ are the relative ratios of the free neutron and free proton densities in the two systems, where $\Delta\mu_n$ and $\Delta\mu_p$ are the differences in the neutron and proton chemical potentials. The empirical observation that this fugacity dependence is respected suggests that the effect of sequential decays on $R_{21}(N,Z)$ is small and that $R_{21}(N,Z)$ reflects the properties of the primary source [2]. If true, $R_{21}(N,Z)$ may be an important and robust observable. Furthermore, Eq. (1) allows one to extrapolate isotope yields over a wide range of the reacting systems from the measurements of a few selected isotopes [7].

Since the Grand Canonical limit is strictly valid only for statistical fragment production in an infinite dilute equilibrated system, it is important to study the validity of the scaling behavior of Eq. (1) with more realistic models. In this paper, we demonstrate that the isoscaling property of Eq. (1) is also predicted by three additional statistical models, the microcanonical and canonical Statistical Multifragmentation Models as well as the Expanding Emission Source (EES) model. In all three of them, isoscaling is affected only slightly by sequential decays, and $\alpha$ and $\beta$ are mainly sensitive to the proton to neutron composition of the emitting source. In a future paper, we will discuss predictions of non-equilibrium transport models such as the Boltzmann-Nordheim-Vlasov [11] and Antisymmetrized Molecular Dynamics models [12].

## II. Microcanonical Statistical Multifragmentation model

To explore the effect of secondary decays on $R_{21}(N,Z)$, we first employ a detailed sequential decay simulation to de-excite primary fragments created in the microcanonical statistical multifragmentation model [13]. Such models have been used successfully to describe fragment multiplicity distributions, charge distributions, mean kinetic energies, and mean transverse energies of the emitted particles from multifragmentation processes [14,15]. However, the most commonly used Statistical Multifragmentation Model (SMM) [16, 17] contains only a schematic treatment of the sequential decays of excited fragments and does not include much of the nuclear structure information needed to describe the secondary decay of hot primary fragments. A new improved sequential decay algorithm [13] has been developed to address the secondary decay problem. Each decay from the initial excited fragment is calculated using tabulated branching ratios when available [18], or by using the Hauser-Feshbach formalism [19], when such information is unavailable. Aside from incorporating empirical information on the binding energies of the nuclei, the new algorithm includes accurate structural information such as the discrete bound states and resonant states for nuclei up to Z=15 [13, 20]. This new sequential decay algorithm is coupled to the SMM code of ref. [21], which was chosen mainly for the ease of incorporating the sequential decays of the primary fragments. This newly modified SMM code is referred as SMM-MSU in this article. The physics results should be similar if other SMM codes are used.

As the primary goal of this article is to understand the general behavior of various models, we will refrain from fitting data by varying model parameters. Instead, we will use previous studies as a guide [22, 23] and choose reasonable and consistent parameters in performing the calculations. We have chosen source sizes corresponding to 75% of the collision systems $^{112}$Sn+$^{112}$Sn and $^{124}$Sn+$^{124}$Sn, an excitation energy of $E^*/A$=6 MeV, and a breakup density of $1/6\rho_0$. The general conclusion of this paper would not change if other source sizes were used. We characterize the neutron and proton composition of the source by the neutron to proton ratio, N/Z or the isospin asymmetry $\delta$=(N-Z)/A=(N/Z-1)/(N/Z+1).

To examine the effects of secondary decay, the predicted carbon isotope distributions from SMM-MSU are shown in Figure 1. The primary distributions from a source of A=186, N/Z=1.48 are shown as open points joined by a dashed line while the final distributions after secondary decay are shown as closed circles joined by a solid line in the top panel. The primary distributions are wide and spread over a large range of neutron-rich nuclei and peak around $^{14}$C. After sequential decays, the distributions are much narrower and peaked near $^{12}$C, more in agreement with experimental observation. Such narrowing of isotope distributions due to sequential decays has been well established [13,24-26].

It has been suggested in Ref. [27] that the isotope distributions are sensitive to the proton and neutron composition of the sources from which the fragments are emitted. To explore this issue, we eliminate the size effect by changing the charge of the emitting source but keeping the size constant, i.e. A=186. The carbon isotope distribution after secondary decay with N/Z=1.48 (closed circles) and N/Z=1.24 (open squares) are compared in the bottom panel of Figure 1. As expected, more neutron rich isotopes (A >12) are produced from the neutron richer system, while the opposite is true for the proton-rich isotope yields. This trend is consistent with experimental observation [2]. It suggests that isotope yield distributions can be used to study properties that reflect the neutron to proton composition of the emitting sources.

Figure 1 illustrates an important point that the isospin effects on isotope yields are much reduced by sequential decays. The differences between the final isotope yields from two systems with different isospin asymmetry are much less than those between primary and final isotope distributions. It is thus important to search for observables such as relative isotope ratios, which cancel out some of the effects of sequential decays, binding energy etc. on isotope productions.

In Figure 2, the relative isotope ratios $R_{21}(N, Z)$ are plotted, as a function of N for the primary and secondary isotope yields predicted by the SMM-MSU model. We choose $A_1$=168 and $Z_1$=75 ($N_1/Z_1$=1.24, $\delta_1$=0.107) and $A_2$=186, $Z_2$=75 ($N_2/Z_2$=1.48, $\delta_2$=0.194) for sources 1 and 2 where $A_i$ and $Z_i$ are the mass and charge number of source **i**. Ratios constructed from primary (final) yields are plotted in the top (bottom) panel. The open symbols represent $R_{21}(N, Z)$ of odd-Z elements while the closed symbols are predicted ratios for the even-Z elements. The ratios of both primary and secondary fragments closely follow the trend described by Eq. (1); isotopes of the same Z, plotted with the same symbol, lie along lines with similar slope in the semi-log plots. For comparison, the solid and dashed lines correspond to the calculations using the best-fit values of $\alpha$, $\beta$ and C of Eq. (1) to the predicted ratios. Since more neutron-rich isotopes are produced from the neutron-rich system, the slopes of these lines are positive.

More importantly, the slopes are similar for all elements before and after sequential decay. This result seems surprising considering the big difference between the primary and secondary distributions shown in the top panel of Figure 1, but it corroborates the assumption that $R_{21}(N, Z)$ is not very sensitive to sequential decays and justifies the empirical approach of Eq. (2) to approximate the effect of sequential decays by a constant multiplicative factor for reactions with similar excitation energy and temperature [2]. The exponential dependence on Z in Eq. (1) suggests that the vertical spacing between adjacent elements should be the same. However this latter requirement is not strictly observed in the predicted results, especially for the final yield ratios. The solid and dashed lines in the upper panel show the best fits of equation 1 with $\alpha$=0.40, $\beta$=-0.50. The scaling parameters extracted after secondary decays in the bottom panel is the same for the neutron slope parameter, $\alpha=0.40$, but the proton slope parameter $\beta=-0.41$ is different, which may indicate the importance of Coulomb effects [28].

For oxygen isotopes, the agreement between predicted ratios after sequential decays and the best fit lines is not very good. This discrepancy may be an artifact from the sequential algorithm used. The current secondary decay code which has structural information for nuclei up to Z=15 may not be reliable for secondary yields with large Z. The effect of incomplete structural information on sequential decays is illustrated in Figure 3. The histograms represent calculations for the carbon (upper panel) and oxygen (lower panel) isotope distributions which use the Hauser Feshbach decay formalism [19] and take into account all the experimental structural information up to Z=15. Closed points joined by dashed lines are the isotope distributions when the Hauser Feshbach formalism is used with the experimental structural information up to Z =10 only [13]. In both cases, decays of heavier fragments not calculated via the Hauser Feshbach approach are calculated with the Weisskopf formalism and liquid-drop binding energies [19]. While the yields for the carbon isotopes are similar with both decay tables, the yields for the neutron rich oxygen isotopes are quite different. Sequential decay calculations with more complete structure information predict more yields for neutron-rich oxygen isotopes. This indicates that sequential charged particle decay plays an important role in producing neutron-rich isotopes and that structure information is relevant to such calculations.

To explore the influence of different sequential decay schemes on isoscaling, the same systems described above ($A_1=168$, $Z_1=75$ and $A_2=186$, $Z_2=75$) are calculated with the more widely used SMM code of Botvina [14-17]. This version of SMM has a simplified description of secondary decay [16,17]; excited light fragments (A<16) undergo fermi breakup while heavier fragments decay by evaporating light nuclei. Figure 4 shows the isotope ratios before and after the sequential decays. The primary yield ratios (upper panel) show the trends as predicted by Eq. (1) but the heavier isotopes (Z≥5) in the final yield ratios (bottom panel) are not as well behaved. This can be attributed to the simplified sequential decay treatments used. The best fit parameters of Eq. (1) are listed in the figure. Predictions from the best fit parameters are plotted as dashed and solid lines. The lines do not describe the predicted ratios after sequential decay well (lower panel). However, the fitted values for α are little altered by sequential decays while the fitted values for β are changed greatly.

**III. Expanding Emitting Source model**

In this section, we examine the Expanding Evaporating Source (EES) model [29] which provides an alternative description of multifragmentation. The EES model utilizes a rate equation formula similar to the evaporation formalism. The emission rate of fragments with $3 \leq Z \leq 20$ is enhanced when the residue expands to sub-saturation density. Within the context of this model, α can be described analytically and provide some physics insight regarding the symmetry energy [7].

Figure 5 shows the relative isotope ratios predicted for multifragmentation processes by the EES model [29] for the systems, $A_1=168$, $Z_1=75$ and $A_2=186$, $Z_2=75$. Even though chemical potentials are not a theoretical ingredient of the EES model, the predicted isotope ratios display isoscaling similar to Equation (1). As in the case of Botvina's SMM calculations, isoscaling is more rigorously observed by the primary yield. Some of the deviations from isoscaling obtained with the final yields may be caused by inaccuracies in the treatment of sequential decays. For example, the EES model includes structural information mainly for low mass nuclei and no information about the unstable particle states for any but the lightest nuclei. Even so, the scaling parameters obtained before and after sequential decays are not very different.

To understand the origin of isoscaling in the EES approach, we must examine the EES fragment emission rate. Similar to the formalism of Friedman and Lynch [30], statistical decay rates in the EES model are derived from detailed balance following the Weisskopf model [31]. When the relative rates are dominated by emission within a particular window of source-mass or source-temperature, the relative yields are directly related to the instantaneous rates

$$dn(N,Z)/dt \propto T^2 \cdot \exp(-V_c/T + N \cdot f_n^*/T + Z \cdot f_p^*/T - B/T) \qquad (3)$$

where $V_c$ gives the Coulomb barrier, and the terms $f_n^*$ and $f_p^*$ represent the excitation contributions to the free energy per neutron and proton, respectively. The factor $B=BE(N_i,Z_i)-$ $BE(N_i-N, Z_i-Z) - BE(N,Z)$ reflects the separation energy associated with the removal of the isotope (N,Z) from the parent nucleus, here denoted by the subscript "i".

When constructing $R_{21}(N,Z)$, some terms, such as the binding energy of the emitted isotope, BE(N,Z), cancel out in the ratio, simplifying the analysis of the dependence of $R_{21}(N,Z)$ on N and Z. To use what remains of the N and Z dependence of the separation energy term B, we expand the differences in the binding energies of the residues with neutron number $N_i-N$ and proton number $Z_i-Z$ in a Taylor series as follows:

$$BE(N_2-N, Z_2-Z) - BE(N_1-N, Z_1-Z) \approx a \cdot N + b \cdot Z + c \cdot N^2 + d \cdot Z^2 + e \cdot N \cdot Z \quad (4)$$

Where a, b, c, d and e are coefficients of the Taylor series. Empirically, the coefficients, c, d, and e of the higher terms in $Z^2$, $N^2$ and ZN are surprisingly small. One can approximate the binding energy difference with the two leading order terms that depend on the difference in the proton and neutron separation energies between the two systems, 1 and 2 i.e. $a=\Delta s_p$, $b=\Delta s_n$. Assuming for simplicity that the residues for systems 1 and 2 have the same charge, $R_{21}(N,Z)$ can be written as follows:

$$R_{21}(N,Z) \propto \exp[\{(-\Delta s_n + \Delta f_n^*) \cdot N + (-\Delta s_p + \Delta f_p^* + e\Delta\Phi(Z_i-Z)) \cdot Z\}/T] \quad (5)$$

where $\Delta\Phi(Z)$ is the difference between electrostatic potential at the surface of residue 1 and residue 2. $\Delta f^*$ is the differences in free energy for the two systems. Aside from the second order term from the electrostatic potential, which is small for the decay of large nuclei, all terms in the exponent of Eq. 5 are proportional to either N or Z, resembling Eq. (1). The corresponding scaling parameters α and β are functions of the separation energies, the Coulomb potential and small contributions from the free excitation energies.

In general, the contribution from free energy is found to be much smaller than the contribution from the separation energy. This is particularly true for systems of comparable mass and energy but different N/Z ratio. Moreover, the volume, surface, and Coulomb contributions to the separation energy largely cancel if the masses of the parent nuclei are similar, leaving the difference in symmetry energies alone as the dominant contribution to $\Delta s_n$. The symmetry energy takes the form:

$$E_{sym} = C_{sym}(N-Z)^2/A = C_{sym}(A-2Z)^2/A \quad (6)$$

The change in neutron separation energy between the two systems can be approximately obtained by taking the derivatives in Eq. (6) with respect to N to obtain

$$\alpha = \Delta s_n / T \approx 4 C_{sym} [(Z_1/A_1)^2 - (Z_2/A_2)^2]/T \tag{7}$$

This dependence leads to a non-linear dependence on $N_i$ and $Z_i$ and a linear relationship between $(Z_2/A_2)^2$ and $\alpha$ for a fixed system 1. In the liquid drop model, $C_{sym}$ takes the value of 23.4 MeV [32]. In the EES model, the symmetry energy term, $C_{sym}$, must be extrapolated to sub-saturation density as the system expands, i.e., $C_{sym}$ is density dependent. Measurements of $R_{21}(N,Z)$ may thus probe the density dependence of the symmetry energy as discussed in Ref. [7].

**IV. Canonical Model**

To explore the relationship between the neutron and proton composition of the source $(Z_2/A_2)$ and $\alpha$ in the statistical fragmentation models, we must perform calculations with different sources. To simplify the discussions, we will use the two fitting parameters, $\alpha$ and $\beta$, which are the average slopes of the lines in the semi-log plot of isotope and isotone yield ratios respectively as shown in Figs. 2, 4 and 5. Since sequential decay does not affect the scaling parameters strongly, we confine our exploration to the influence of the parameters on the primary distributions.

For these studies we use the statistical multifragmentation model (SMM-McGill) [27] that uses recursive techniques to shorten the time needed for a canonical calculation. We have compared the predictions of this canonical approach to the microcanical model of ref. [13]; the two approaches provide similar predictions for the observables presented below. There is also a similarity between both approaches and the predictions of the Grand Canonical ensemble [26, 33].

Canonical model predictions for the temperature and density dependences of $\alpha$ are shown in the left and right panels of Figure 6, respectively. The calculations assume a fixed freeze-out density of $\rho_0/3$ in the left panel, and fixed temperatures of 4, 5 and 6 MeV in the right panel. The same systems, $A_1=168$, $Z_1=75$ and $A_2=186$, $Z_2=75$, are used. Isospin effects decrease with increasing T. There is a significant sensitivity to temperature at low temperature, but both the sensitivity to temperature and the overall isospin effect diminish at very high temperature. On the other hand, $\alpha$ is less sensitive to the breakup density.

It is interesting to note that if one were able to constrain the temperature and density with experimental information, the connection between α and the N/Z ratio of the fragmenting system could be used to constrain the latter quantity. This sensitivity is useful to constrain the N/Z of the fragmentating sub-system (prefragment) if it is modified by the preequilibrium emission prior to breakup. Transport calculations predict that the relative neutron vs. proton preequilibrium emission may be sensitive to the density dependence of the asymmetry term of the nuclear equation of state [34]. If so, charge and mass conservation implies that observables sensitive to the N/Z of the prefragment may provide constraints on the density dependence of this asymmetry term [20].

The temperature dependence of the difference in chemical potentials, $\Delta\mu_n = \alpha \cdot T$ and $\Delta\mu_p = \beta \cdot T$, is shown in the left and right panel of Figure 7, respectively. If the change in the chemical potentials for the two systems as a function of temperature were the same, then $\Delta\mu_n$ and $\Delta\mu_p$ would be constant. Instead, we see a decrease in the differences between the chemical potential, with increasing temperature. Interestingly, there is a break in the slope at T=5 MeV. There is currently no satisfactory explanation for such a break. Further studies are needed.

Experimentally, a nearly linear relation between the "relative free neutron density", $\hat{\rho}_n = \exp(\alpha)$ and $(N_2/Z_2)$ ratio of the system 2, has been observed [2] over the range of $(N_2/Z_2)$ from 1.24 to 1.48. To explore this issue within the context of Eq. (7), we kept our reference system (reaction 1) fixed at $A_1=168$, $Z_1=75$ and performed calculations on systems with different $(N_2/Z_2)$ values. The results are shown in the right panel of Figure 8. Four groups of calculations are performed by either keeping source size constant at $A_2=186$ (solid circles), or $A_2=124$ (open circles) or by keeping the charge of the source constant at $Z_2=75$ (closed squares) or $Z_2=50$ (open squares). In all cases, the slope parameters α are dependent mainly on $N_2/Z_2$ or equivalently on the isospin asymmetry $\delta_2=(N_2/Z_2-1)/(N_2/Z_2+1)$, of system two and independent of its charge number and source size. The experimental linear relationship between $\hat{\rho}_n$ and $(N_2/Z_2)$ is observed only within a narrow range of $(N_2/Z_2)$ from 1.2 to 1.5 as demonstrated in the right panel of Figure 8. Over a larger range of $(N_2/Z_2)$, there is a concave curvature in $\hat{\rho}_n$ which is especially noticeable at small $N_2/Z_2 \leq 1$. The relationship between α and $N_2/Z_2$ in the right panel is best described by $\alpha=3.0-15.21/(1+N_2/Z_2)^2$ (solid line).

The SMM model describes an instantaneous multifragmentation process rather than a sequential binary breakup process. Interestingly, the linear relationship between $\alpha$ and $(Z_2/A_2)^2$ predicted by the EES model in Eq. (7), is also evident in the SMM calculations as shown in left panel of figure 8. If T is taken to be 5 MeV, we obtain $C_{sym}$=19.2 MeV as compared to the liquid drop value of 23.4 MeV. Such relationship is perhaps not so surprising at low excitation energy, where recent SMM model calculations [13] indicate that $\mu_n$ and $s_n$ are closely related as expected. At high excitation energy, the role of multifragment decay configurations become important. There is no direct connection between $\mu_n$ to $s_n$ a priori. It is thus intriguing to see that the linear relationship of Eq. (1) is preserved and that Eq. (7) is valid even at high excitation energies. Such dependence probably signals the importance of the symmetry energy as the dominant contribution to $\alpha$ in the SMM model. Indeed, if the symmetry terms to the binding energies of the nuclei are turned off in the SMM and EES calculations, the isoscaling behavior observed in Figure 2, 4 and 5 will disappear.

**V. Summary**

We have calculated the isotope distributions from Z=1 to Z=8 particles using different multifragmentation models. The simple factorization of $R_{21}(N,Z)$ into the neutron and proton "fugacity" terms has been demonstrated by all the models studied in this article. The relative isotope ratios are not affected very much by the sequential decays, so in these statistical models $R_{21}(N,Z)$ reflects the isotope yield ratios of the primary fragments. The isotope distributions are determined mainly by the isospin asymmetry of the emitting source and to a lesser extent the temperature of the system. For statistical models, it appears that $R_{21}(N,Z)$ provides an opportunity to study isotopic observables that are related to the primary fragmentation process. This may provide access to the early stages of the fragmentation process where there may be sensitivity to the symmetry terms of the equation of state, which directly influence the neutron to proton ratios of the intermediate emission source.

This work is supported by the National Science Foundation under Grant No. PHY-95-28844, PHY-96-05140, INT-9908727 and contract No. 41.96.0886.00 of MCT/FINEP/CNPq (PRONEX).

**FIGURE CAPTIONS:**

Figure: Differential multiplicities at $\theta_{CM}=90°$ for carbon isotopes as a function of the mass number of the isotope. Top panel: primary yields are denoted by open points connected by the dashed lines while the solid points joined by solid lines denote the yield after sequential decays (see text for details). Bottom panel: Carbon isotope yields for two systems with different isospin asymmetries, closed circles for $\delta=0.194$, N/Z=1.48 and open squares for $\delta=0.107$, N/Z=1.24. The source size is kept constant at A=N+Z=186.

Figure 2: Predicted (symbols) relative isotope ratios, $R_{21}(N,Z)$, of Eq. 1 for the two systems, $A_1=168$, $Z_1=75$ and $A_2=186$, $Z_2=75$ using the SMM_MSU code [13,20] as a function of N obtained from the primary isotope yields (upper panel) and the final yields after sequential decays (lower panel). Solid and dashed lines are best fits to Equation 1 using the predicted ratios.

Figure 3: Differential multiplicities at $\theta_{CM}=90°$ for carbon (top panel) and oxygen (bottom panel) isotopes as a function of the mass number. Closed points are predictions if the sequential decay information from Ref. [13] where the sequential decay table truncates at Z=13, is used. Histograms are predictions when the structure information in the structural information table of Ref [13] is extended to Z=15 [20].

Figure 4: Predicted (symbols) relative isotope ratios, $R_{21}(N,Z)$, for the same systems as in Figure 2, using the SMM code of Ref. [17].

Figure 5: Predicted (symbols) relative isotope ratios, $R_{21}(N,Z)$, for the same systems as in Figure 2, using the EES code of Ref. [29].

Figure 6: Temperature (left panel) and density (right panel) dependence of the scaling parameter, α. The sources used are the same as those in Figure 2.

Figure 7: Temperature dependence of the difference in neutron (left panel) and proton (right panel) chemical potentials obtained from the canonical SMM calculations.

Figure 8: α as a function of $(A_2/Z_2)^2$ for four calculations with constant source size (solid and open circles) or constant charge (solid and open squares). The linear relationship shown in the left panel follows Eq. (7). In the right panel, the relative free n-density, $\hat{\rho}_n$, is plotted as a function of the $N_2/Z_2$ ratio of the source. A linear relationship is observed over the range of $N_2/Z_2$=1.24 and 1.48, similar to the experimental results. However, over a wider range, the dependence of $\hat{\rho}_n$ on $N_2/Z_2$ is far from linear.

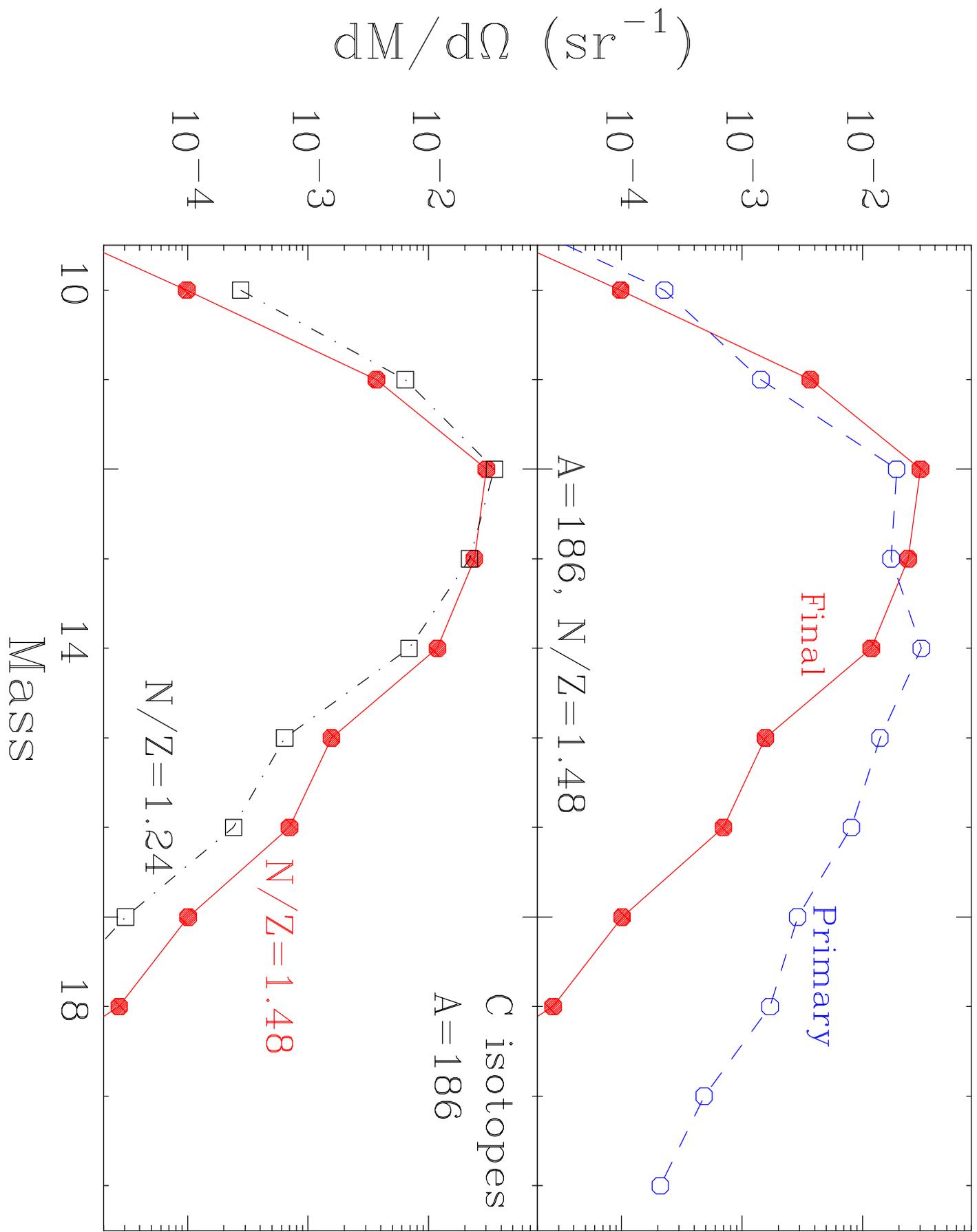

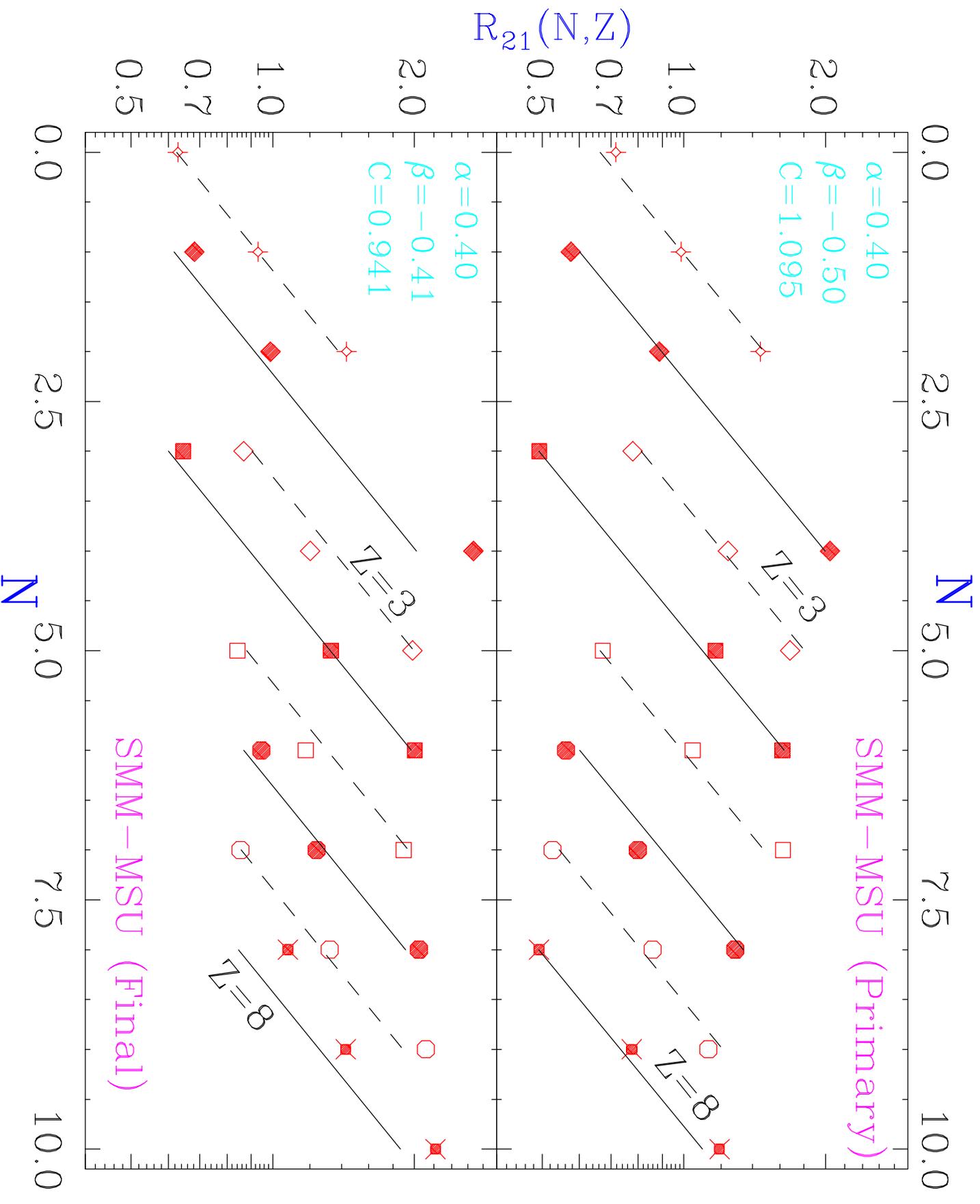

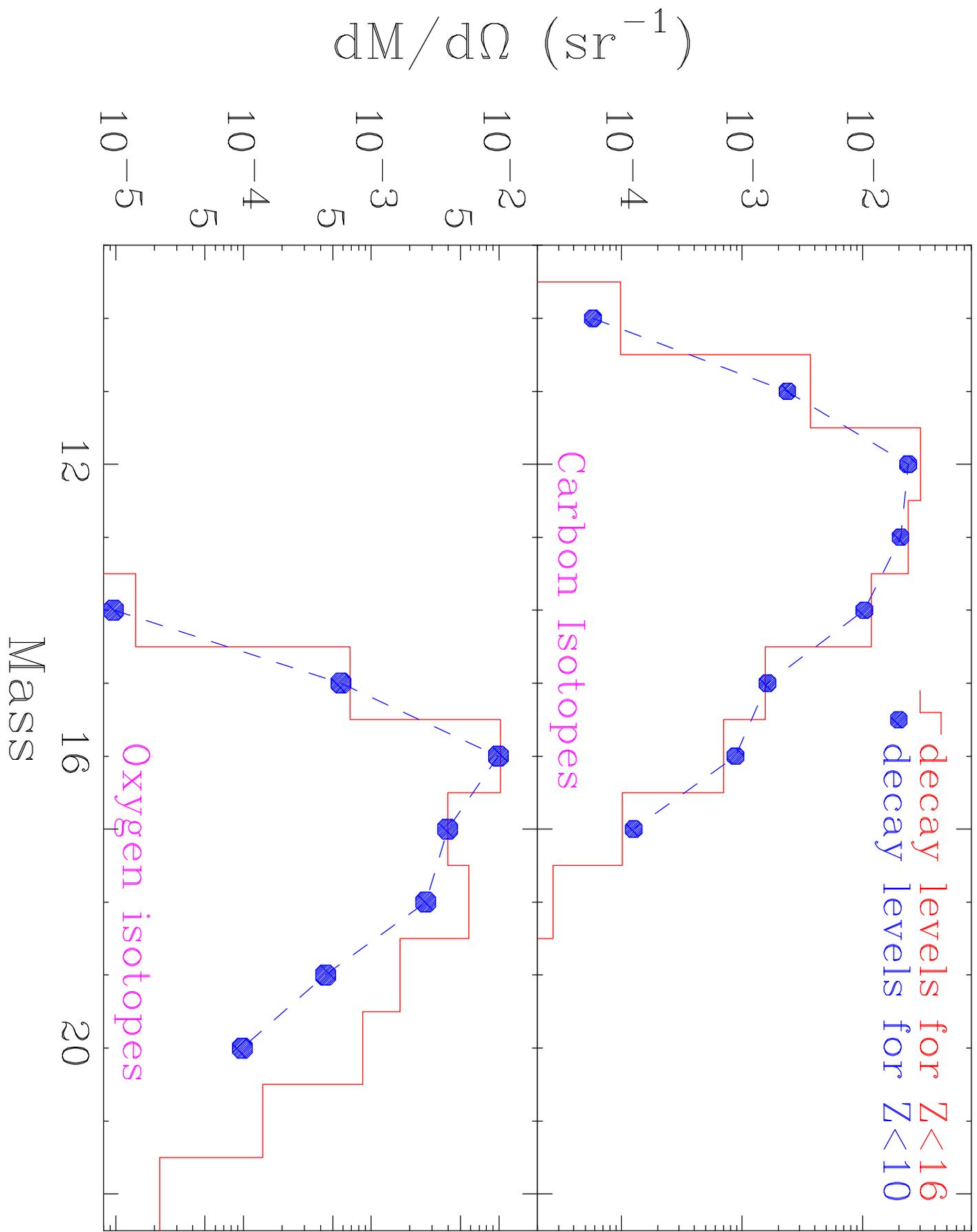

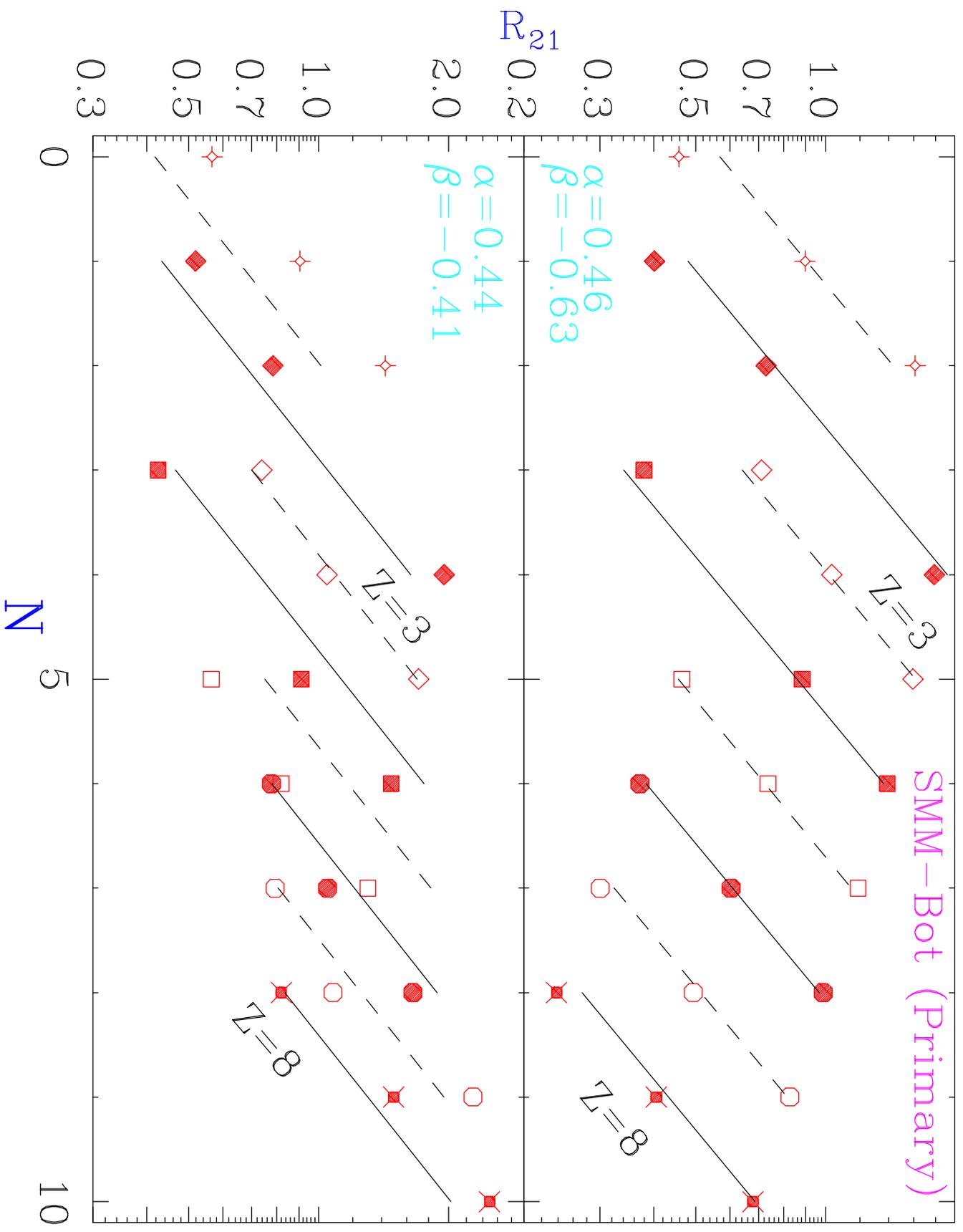

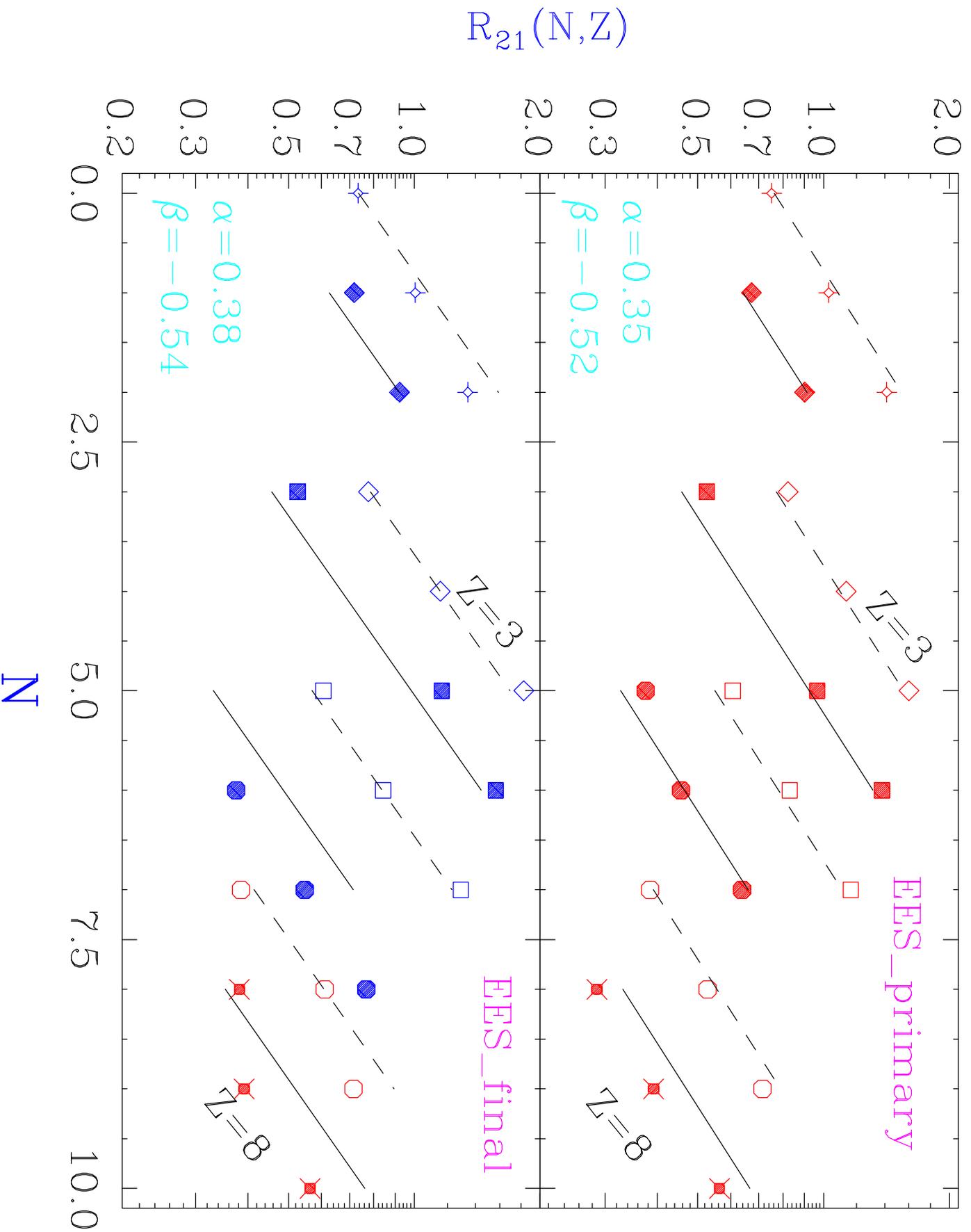

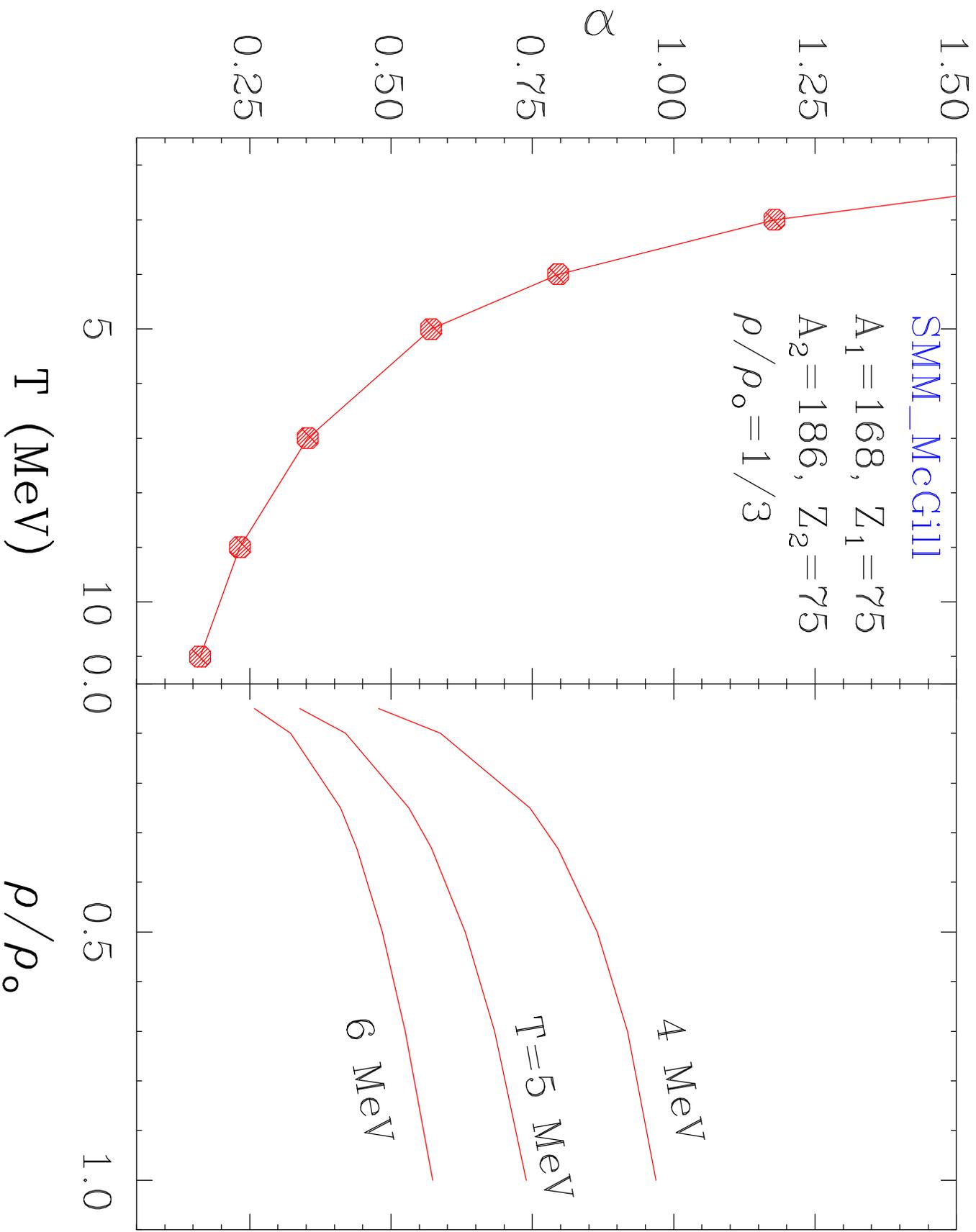

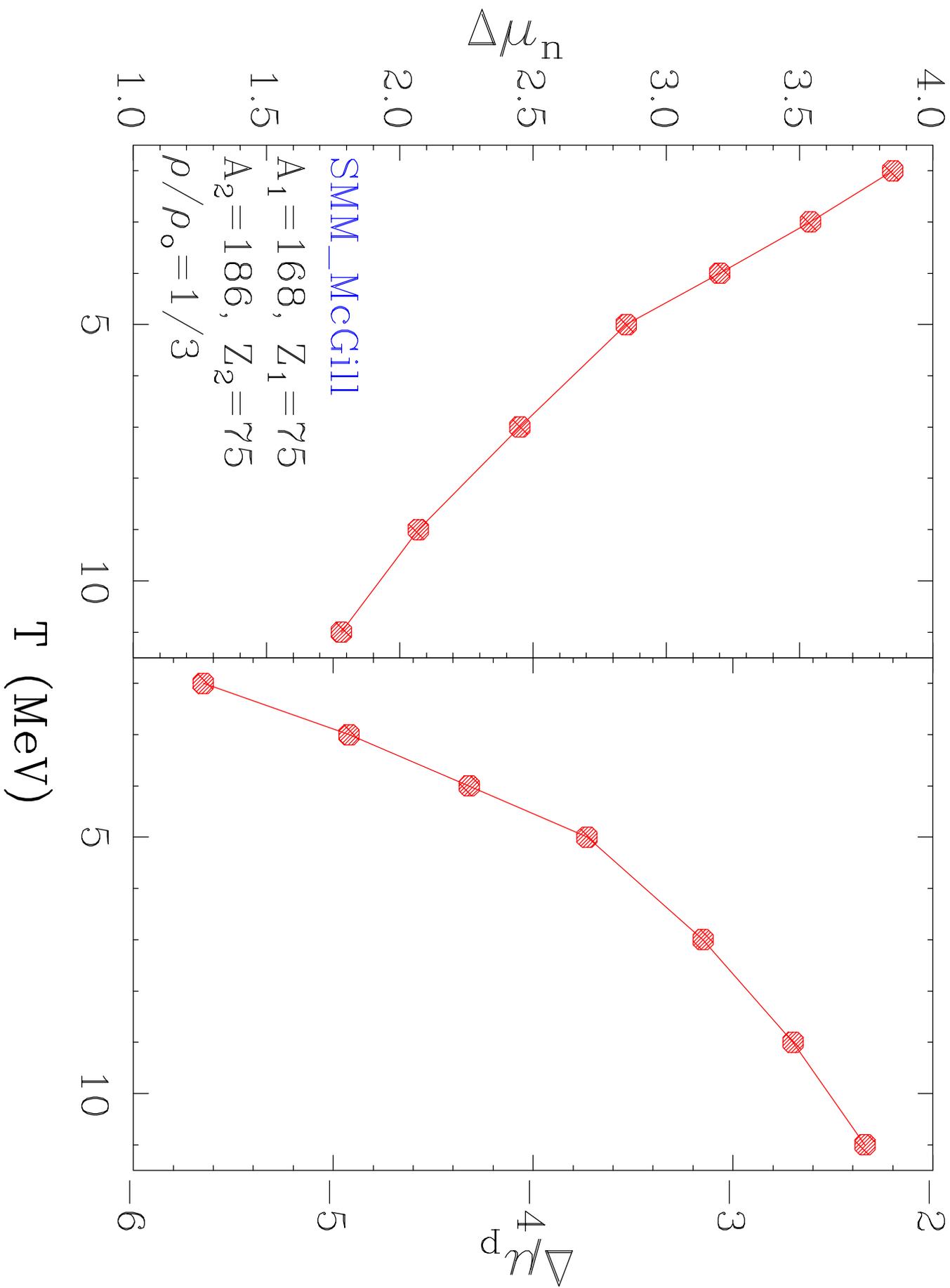

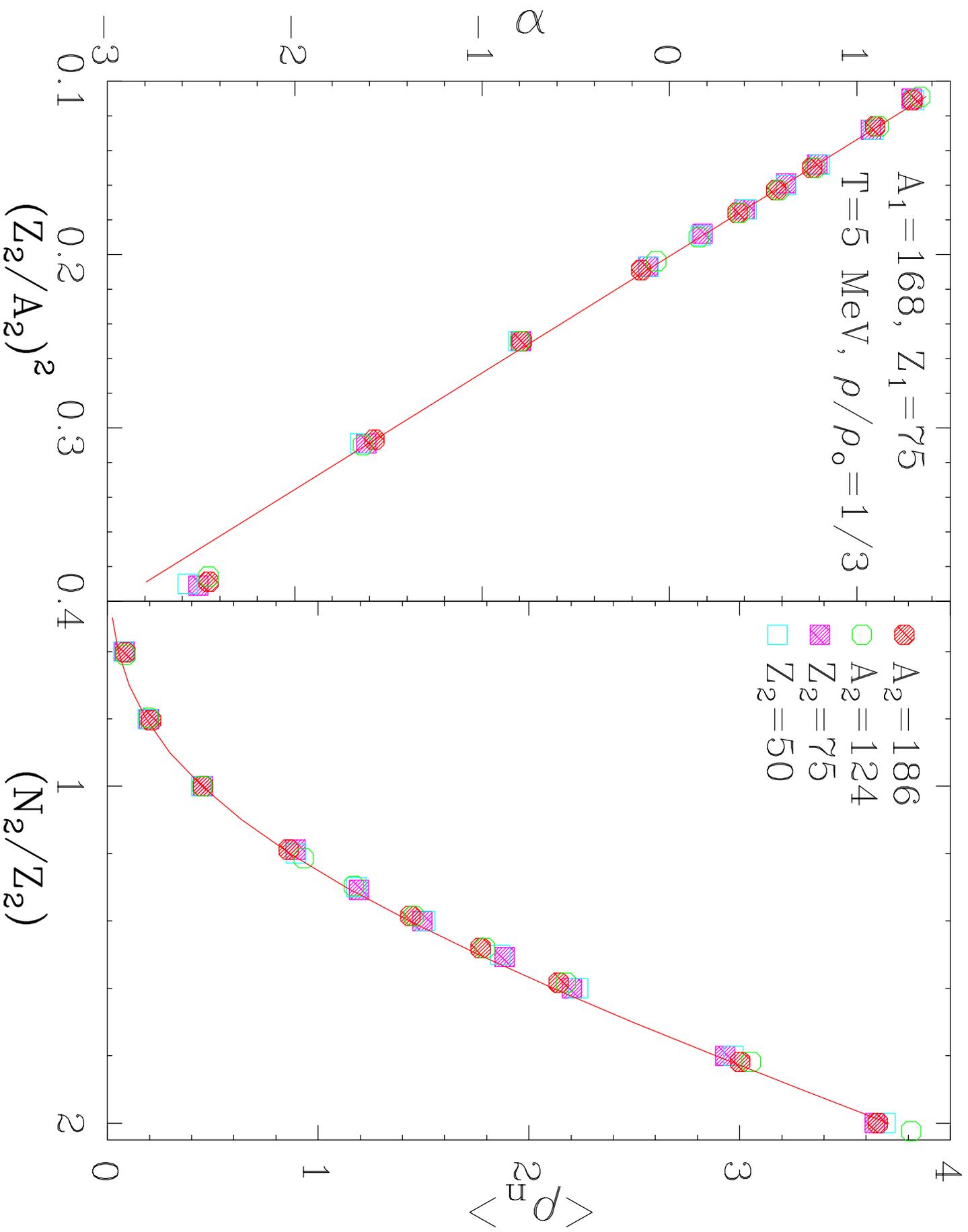